%% file: main_camerReady.tex
\documentclass[conference]{IEEEtran}
\IEEEoverridecommandlockouts
\usepackage{cite}
\usepackage{amsmath,amssymb,amsfonts}
\usepackage{algorithmic}
\usepackage{graphicx}
\usepackage{textcomp}
\usepackage{xcolor}
\usepackage{textpos}
\def\BibTeX{{\rm B\kern-.05em{\sc i\kern-.025em b}\kern-.08em
    T\kern-.1667em\lower.7ex\hbox{E}\kern-.125emX}}
\usepackage{multirow,makecell}
\begin{document}
\normalsize

\title{Unifying Model and Layer Fusion for Speech Foundation Models}


\author{
\IEEEauthorblockN{Yi-Jen Shih, David Harwath}
\IEEEauthorblockA{The University of Texas at Austin \\
\texttt{\{yjshih, harwath\}@utexas.edu}
}
}



\maketitle

\begin{abstract}
Speech Foundation Models have gained significant attention recently. Prior works have shown that the fusion of representations from multiple layers of the same model or the fusion of multiple models can improve performance on downstream tasks. We unify these two fusion strategies by proposing an interface module that enables fusion across multiple upstream speech models while integrating information across their layers. We conduct extensive experiments on different self-supervised and supervised models across various speech tasks, including ASR and paralinguistic analysis, and demonstrate that our method outperforms prior fusion approaches. We further analyze its scalability concerning model size and count, highlighting the importance of selecting appropriate upstream models. Our results show that the proposed interface provides an additional performance boost when given a suitable upstream model selection, making it a promising approach for utilizing Speech Foundation Models.
\end{abstract}


\begin{textblock*}{\textwidth}(0cm,13cm)
\tiny
\noindent
\copyright~2025 IEEE.  Personal use of this material is permitted. Permission from IEEE must be obtained for all other uses, in any current or future media, including reprinting/republishing this material for advertising or promotional purposes, creating new collective works, for resale or redistribution to servers or lists, or reuse of any copyrighted component of this work in other works.
\end{textblock*}

\section{Introduction}
\input{sections/introduction}

\section{Related Work}
\input{sections/related_work}

\section{Unifying Model and Layer Fusion}
\input{sections/interface_definition}

\input{tables/main_results_singleCol}
\input{tables/largeModel_results}
\vspace{-10pt}
\section{Experimental Setup}
\input{sections/eval_setup}

\section{Results and Discussion}
\input{sections/discussion}

\section{Detailed Results for different Languages on ML-SUPERB Mono-1h}
\input{sections/discussion_diff_langs}

\section{Conclusion}
\input{sections/conclusion}'

\section*{Acknowledgment}
\small
This material is based upon work supported by the National Science Foundation under Grant Number 2238605. Any opinions, findings, and conclusions or recommendations expressed in this material are those of the author(s) and do not necessarily reflect the views of the National Science Foundation.


\bibliographystyle{IEEEtran}
\bibliography{mybib}

\end{document}

%% file: sections/introduction.tex
Speech Foundation Models (SFMs) have become extremely popular in recent years.
These models are pretrained using either self-supervised learning~(SSL)~\cite{hsu2021hubert,Baevski_wav2vec2,Chen2021WavLMLS,baevski22_data2vec} or a supervised learning~(SL)~\cite{radford23_whisper} objective on large scale training data.
A major appeal of SFMs is that after initial large-scale training, they can be transferred to new downstream tasks with a relatively modest amount of new training data.
Prior literature has demonstrated the generalizability of SFM representations across various downstream speech processing tasks~\cite{yang_SfmEval,shi23g_mlsuperb}, including Automatic Speech Recognition (ASR)~\cite{baevski2021wav2vec-u,liu2022wav2vec-u2}, Speech Segmentation~\cite{peng2022vghubert,Strgar_2022}
Visually Grounded Speech~\cite{shihSpeechCLIP,layne_mspchclip}, Speaker Verification~\cite{aldeneh24_SVDS} and Emotion Recognition~\cite{Morais_emotion_ssl}.


Despite their strong representations, researchers have explored methods to further enhance the effectiveness of SFMs.
One such approach involves fusing multiple speech encoders to obtain more powerful representations for downstream tasks. 
This concept dates back to the 1990s as ``ensemble learning''~\cite{Hansen90_ensemble,schapire90_strength,Breiman96_BaggingP,Freund_adaboost} whereby weak classifiers can be combined into a strong classifier.
Ensembling models can improve generalization ability by leveraging diverse training data distributions and model architectures.
Since SFMs are pretrained using different objectives and different data distributions (varying in acoustic conditions, speaker distributions, and content domains), it is reasonable to combine multiple SFMs to boost performance and robustness.
Several works have shown the benefits of SFM fusion in tasks such as ASR~\cite{tang22_fusion,fu23b_otf}, MOS Prediction~\cite{yang22o_mosFusion}, and Emotion Recognition~\cite{morais22_erFusion}.

Another direction for improving SFMs is optimizing the approach of fusing information across speech encoder layers.
As observed in~\cite{Ankita2022CCA}, different layers of self-supervised speech models tend to encode different aspects of speech information in an input utterance. 
In general, lower layer representations are highly correlated with the acoustic waveform itself, while deeper layers encode more abstract representations such as phonetic identity.
To leverage SSL models on many different downstream tasks,
a commonly used approach (such as in the SUPERB~\cite{yang21c_superb} benchmark) is to compute a weighted sum over all layer representations (with trainable weights) from the upstream SSL model which is then used as the input to the downstream model. 
In doing so, the extracted information from the SSL model is not limited to a single layer, and the choice of layer features to use can be trained jointly with the task-specific, downstream model.
Prior work~\cite{yang_sfm_eval} has shown that extracting features using weighted-sum outperform selecting single layer in most of the downstream tasks.
While the weighted sum fusion approach is simple and parameter efficient, many recent works~\cite{huo23b_gaff,chiu24_sinica,shih24_interface} have proposed alternative layer fusion approaches that attempt to improve over the weighted sum.
Among them, ~\cite{shih24_interface} introduced the ``Interface'' concept, a learnable module connecting upstream and downstream models. 
Their proposed Hierarchical Convolution (HConv) interface consistently outperformed a learnable weighted sum across multiple ASR and non-ASR tasks and different SSL models.

\input{figs/fusion_overview_1}
In this work, we argue for a unified framework that jointly considers model fusion and layer fusion. 
To the best of our knowledge,~\cite{chiu24_sinica} is the first attempt to jointly address model fusion and layer selection, but their approach still treats the two processes separately by independently performing layer fusion within each model before applying model fusion. 
Moreover, \cite{chiu24_sinica} finds that there is no single combination of model and layer fusion that consistently performs well regardless of the chosen upstream model.
Also, their evaluation is limited only to the ASR task on LibriSpeech using self-supervised foundation models.

Our work addresses these limitations in several ways. 
First, we unify model fusion and layer fusion into a single module and demonstrate that optimizing them jointly significantly outperforms the method proposed by \cite{chiu24_sinica}. 
Second, our evaluation spans multiple ASR and non-ASR tasks, including Emotion Recognition and Speaker Verification, and we show consistent improvements across these tasks using our proposed method. 
Finally, we consider the fusion of both supervised and self-supervised speech models and find that their fusion outperforms that of fusing solely self-supervised models.


In conclusion, our contributions can be summarized as follows:
\begin{enumerate}
    \item We introduce a novel technique for unifying and jointly optimizing model and layer fusion, and show that it outperforms prior work which optimizes these separately.
    \item Our work is the first to conduct thorough experiments on SFM fusion across both ASR and non-ASR tasks.
    \item In most cases, our proposed fusion method outperforms prior methods for various upstream models. We also observe the best performance when fusing self-supervised with traditional supervised speech models, indicating that these models learn to model different and complementary aspects of the speech signal.
\end{enumerate}


%% file: figs/fusion_overview_1.tex
\begin{figure*}[t]
  \centering
   \includegraphics[width=\linewidth,trim={0.8cm 0.3cm 2.2cm 0.2cm},clip]{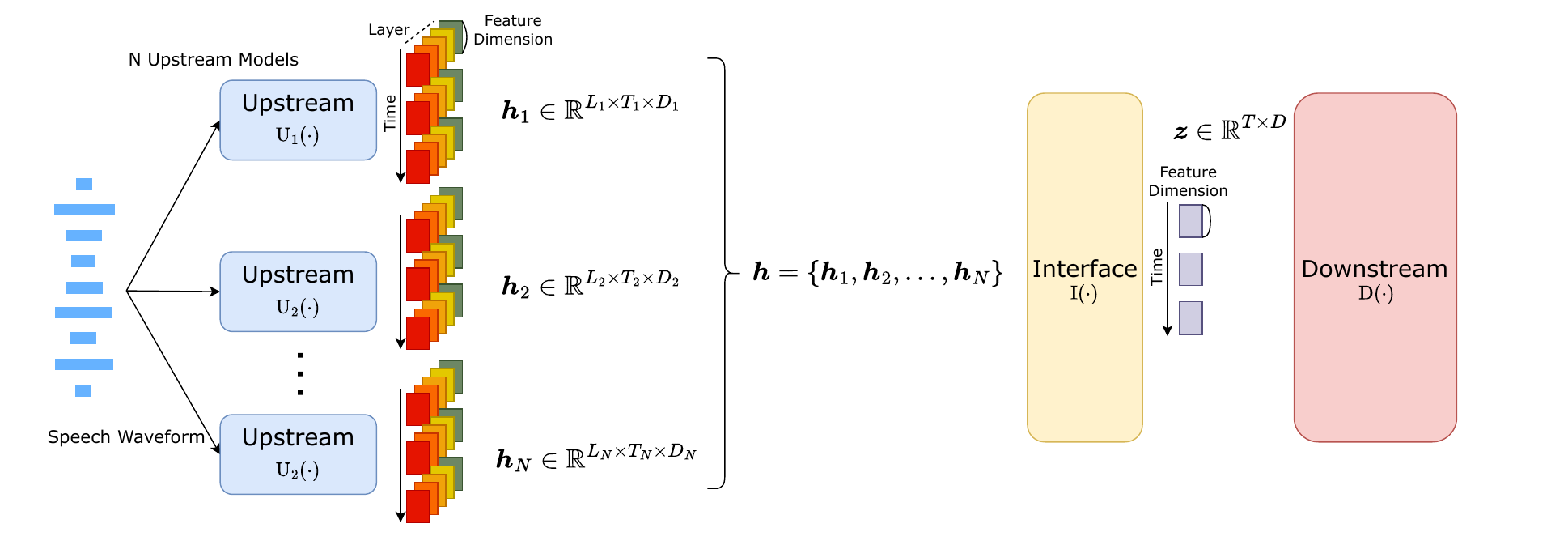}
   \vspace{-20pt}
  \caption{An overview framework of Interface with fusion.}
  \label{fig:overview}
\vspace{-15pt}
\end{figure*}

%% file: sections/related_work.tex
Arising from the concept of ensemble learning in the 1990s, recent works have explored ensemble techniques for Speech Foundation Models (SFMs). \cite{arunkumar22b_SSLFusion} was an early work studying the effect of fusion for self-supervised speech models. 
However, their work only considers performing fusion on the last layer output of each model.
\cite{tang22_fusion} further improves it by extracting multiple layer features from each model while doing fusing.
In~\cite{fu23b_otf}, instead of fusing the layer outputs of multiple SFMs, the authors propose to use Optimal Transport to perform the fusion of model weights directly.
While this method significantly reduces computational cost at inference time, it requires extensive end-to-end fine-tuning on the merged model to achieve competitive performance.
Due to computational constraints, we do not include this approach in our study.
Effuse~\cite{srivastava24_effuse}, on the other hand, trains an auxiliary prediction head to use features from one SFM to predict the output of another SFM, eliminating the need to run inference on both models at testing time.
However, their fine-tuning process consists of two stages: first training the upstream models separately, followed by training the predictor.
Due to this multi-stage fine-tuning requirement, we exclude this method from our experiments.
Nevertheless, our baseline method of using weighted sum can be considered to be equivalent to the first stage of Effuse.
Finally, many of these prior works focus exclusively on ASR tasks, while our work also evaluates speaker-related and paralinguistic tasks.

Several papers have also studied layer fusion for SSL speech models. In~\cite{huo23_hff}, the authors proposed methods to fuse layers in a hierarchical way and~\cite{huo23b_gaff} propose a global attentional feature fusion.
More recently,~\cite{chiu24_sinica,shih24_interface} systematically investigated multiple layer fusion strategies for utilizing Speech SSL models.
The former focused on exploring various combinations of layer and model fusion, whereas the latter concentrated specifically on layer fusion.
In terms of evaluation, the former evaluates only on LibriSpeech ASR, while the latter evaluates on a variety of tasks from SUPERB and ML-SUPERB.


%% file: sections/interface_definition.tex
\subsection{Background on Interfaces for SSL Models}
In~\cite{shih24_interface}, the authors define a framework for utilizing SFM models by specifying 3 modules: the Upstream~($\mathrm{U}$) model, Interface~($\mathrm{I}$), and Downstream~($\mathrm{D}$) model.
The three components implement the following mappings:
\begin{align}
    \mathrm{U}\left(\cdot \right) & : \mathbb{R}^{T'} \mapsto \mathbb{R}^{L\times T\times D} \\
    \mathrm{I}\left(\cdot \right) & : \mathbb{R}^{L\times T\times D} \mapsto \mathbb{R}^{ T\times D} 
\end{align} where $T'$ is the temporal length of an input speech utterance, $T$ is the length in frames after downsampling by the upstream model, $L$ is the number of layers and $D$ is the hidden dimension of the upstream model.
The downstream model takes the output of $\mathrm{I}\left(\cdot \right)$ as input for specific downstream tasks.
For the default setting of this framework, the upstream model is initialized from a pretrained speech encoder and remains frozen throughout the downstream task finetuning, while the interface and downstream are trainable\footnote{In their work, they also try finetuning all components. However, we do not include this setting in our work due to computational constraints.}.
The function of the interface is to aggregate information across all layers of the upstream model, and a good interface should generalize to many downstream tasks while maintaining a modest number of trainable parameters.
\cite{shih24_interface} empirically demonstrated that the widely-used weighted sum interface is susceptible to information collision across layers, making it suboptimal.
To this end,~\cite{shih24_interface} proposed an alternative interface design (HConv), which uses a series of 1D convolution layers applied across the upstream model's layer dimension. HConv demonstrated significant improvements over the weighted sum interface across various speech processing tasks with different upstream models.

\subsection{Extending Interfaces to Model and Layer Fusion}
In this work, we extend the definition of the Interface framework to encapsulate not only fusion across layers, but also fusion across multiple upstream models.
Formally, suppose we have $N$ Speech Foundation Models $\left[\mathrm{U_1},\mathrm{U_2}\dots \mathrm{U_N}\right]$ as our upstream models.
After feeding the input utterance $\mathbf{x}\in \mathbb{R}^{T'}$ into each upstream model separately, we obtain a list of 3-dimensional tensors by extracting all hidden representations of each model. i.e., $\mathbf{h}=\left\{\mathbf{h}_1,\mathbf{h}_2,\dots,\mathbf{h}_N\right\}$.
Each hiddenstate $\mathbf{h}_i\in \mathbb{R}^{ L_i \times T_i \times D_i}$ is allowed to have different numbers of layers, temporal length, and hidden dimension. 
Under this relaxation of the framework, an interface module jointly performs model and layer fusion by implementing the mapping $\mathrm{I}\left(\cdot \right): \left( 
\mathbb{R}^{L_1\times T_1\times D_1},
\mathbb{R}^{L_2\times T_2\times D_2},
\dots,
\mathbb{R}^{L_N\times T_N\times D_N}
\right) \mapsto \mathbb{R}^{ T\times D}$.


To allow fusion for multiple upstream models, we revise the Hierarchical Convolution Interface.
First, we must ensure that all $L_i$ and $T_i$ are identical across models. i.e. $\forall i,j, 1\leq i,j \leq N, L_i=L_j \wedge T_i=T_j$.
While most of the commonly used Speech Foundation Models utilize a framerate of 50Hz and typically use the same number of layers (12 Transformer layers for base models and 24 for Large models), in the case that the number of layers or the framerates differ for some of the upstream models, we upsample the representations of the models to match that of the model with the highest framerate (and similarly with the number of layers), and then apply linear interpolation to both axes.
We denote the matched layer and temporal dimension as $L$ and $T$ respectively.
After matching $L,T$, we have two options for merging the hidden states of each model.
If the hidden dimensions of each model are the same (or interpolation is applied to make it the same), we can directly add all models' hidden state together to obtain a tensor as $\mathbb{R}^{L \times T\times D}$.
The other option is to directly concatenate all models' hidden states to obtain a tensor as $\mathbb{R}^{L \times T\times \left(N \cdot D\right)}$.
We can then insert the merged tensor into the Hierarchical Convolution Interface.
We call the former method ``HConv.'' and latter ``CHConv.''
Note that the merging operations mentioned above possess the characteristic of local similarity on the last dimension of the tensor, which still complies with the assumption of using the Hierarchical Convolution Interface.


%% file: tables/main_results_singleCol.tex
\begin{table}[h!]
    \setlength{\tabcolsep}{4pt}
  \caption{Evaluation results for different interfaces across ASR and non-ASR tasks. ``LS'' denotes LibriSpeech, ``Mono'' and ``Multi'' indicates Mono-1h and Multi-1h on ML-SUPERB. Their metrics are WER/CER. ``SV'' and ``ER'' stands for Speaker Verification and Emotion Recognition respectively and their metrics are Equal Error Rate and Accuracy. For the upstream models, ``Hbrt'', ``WvLM'', ``Whspr'' denotes ``HuBERT Base'', ``WavLM Base+'' and ``Whisper Small''. 
  The interfaces are abbreviated as the following: ``WS.'' : ``Weighted Sum'', ``HConv.'': ``Hierarchical Convolution'' and ``GumD.'': ``Gumbel Dimension Selection''.
  The best scores within each each section are underscored and the best score in general is bolded. 
  }
  \label{tab:main_result_single_col}
\small
  \centering
  \begin{tabular}{lcccccc}
    \hline
    
      \multirowcell{ 2}{\textbf{Model}} & \multirowcell{ 2}{\textbf{Interface} } & \multicolumn{3}{c}{\textbf{ASR}}  & \multicolumn{2}{c}{\textbf{Non-ASR}} \\
      \cline{3-7}
          &   & \textbf{LS}~$\downarrow$ & \textbf{Mono}~$\downarrow$ & \textbf{Multi}~$\downarrow$  & \textbf{SV}~$\downarrow$ & \textbf{ER}~$\uparrow$   \\
    \hline
  \hline
    \multicolumn{7}{c}{\textit{Single Model}} \\
    \hline

\multirowcell{ 2}{Hbrt} 
& WS.~\cite{yang21c_superb}			& 6.32           & 34.66 & \underline{32.6} & 3.93 & 65.71 \\
& HConv.		& \underline{5.80}  & \underline{33.66} & 32.9 & \underline{3.63} & \underline{69.78} \\
\hline
\multirowcell{ 2}{WvLM} 
& WS.~\cite{yang21c_superb}			& 5.40  &  32.01 & 	29.2 & 3.52 & 66.95 \\
& HConv.		& \underline{4.78}  &  \underline{30.74} & 	\underline{27.6} & \underline{2.79} & \underline{70.76}  \\
\hline
\multirowcell{ 2}{D2v} 
& WS.~\cite{yang21c_superb}			& 4.83 & 	 	35.77 & 	34.2 & 4.87 & 65.60  \\
& HConv.		& \underline{4.53}  & 	\underline{34.22} & 	\underline{33.5} & \underline{4.20} & \underline{69.33}  \\
\hline
\multirowcell{ 2}{Whspr} 
& WS.~\cite{yang21c_superb}			& 6.73  &  	26.27 & 	23.8 & 4.64 & 69.33 \\
& HConv.		& \underline{6.40}  & 	\underline{26.23} & 	\underline{21.7} & \underline{3.30} & \underline{70.36}  \\
\hline
\hline
\multicolumn{7}{c}{\textit{2 Model Fusion}} \\
\hline
\multirowcell{ 4}{Hbrt  \\ \& WvLM } 
& WS.~\cite{chiu24_sinica}			& 5.42 & 31.89 & 29.4 & 3.59 & 68.29 \\
& GumD.~\cite{chiu24_sinica}	        & 5.89 & 32.83 & 30.3 & 5.66 & 61.83 \\
& HConv.		& 4.86 & 30.65 & 31.8 & \underline{\textbf{2.74}} & 70.49 \\
& CHConv.		& \underline{4.82} & \underline{30.40} & \underline{29.8} & 2.88 & \underline{70.55} \\
\hline
\multirowcell{ 4}{D2v  \\ \& WvLM }
& WS.~\cite{chiu24_sinica}			& 4.76 & 32.02 & \underline{29.1} & 3.56 & 68.29 \\
& GumD.~\cite{chiu24_sinica}     	& 4.72 & 32.93 & 30.2 & 6.37 & 62.07 \\
& HConv.		& \underline{\textbf{4.30}} & \underline{30.62} & 30.5 & \underline{2.91} & \underline{71.95} \\
& CHConv.		& 4.39 &  30.82 & 31.1 & 3.06 & 70.93 \\
\hline
\multirowcell{ 4}{D2v  \\ \& Hbrt }
& WS.~\cite{chiu24_sinica}			& 4.87 & 33.80 & \underline{32.1} & 4.01 & 66.98 \\
& GumD.~\cite{chiu24_sinica}	        & 4.94 & 34.84 & 32.8 & 6.25 & 61.88 \\
& HConv.		& 4.60 & \underline{32.29} & 33.3 & 3.92 & \underline{70.80} \\
& CHConv.		& \underline{4.47} & 32.40 & 32.5 & \underline{3.72} & 69.94 \\


\hline
\multirowcell{ 4}{Whspr  \\ \& Best SSL }
& WS.~\cite{chiu24_sinica}			& 4.77 & 25.89 & 22.9 & 3.96 & 70.58 \\
& GumD.~\cite{chiu24_sinica}	        & 4.70 & 27.37 & 23.3 & 5.95 & 65.24 \\
& HConv.		& 4.93 & 24.88 & 21.1 & \underline{2.90} & 71.29 \\
& CHConv.		& \underline{4.52} & \underline{\textbf{23.54}} & \underline{\textbf{20.2}} & 3.03 & \underline{\textbf{74.86}} \\

\hline
\hline
\multicolumn{7}{c}{\textit{3 Model Fusion}} \\
\hline

\multirowcell{ 3}{D2v  \\ \& Hbrt \\ \& WvLM } 
& WS.~\cite{chiu24_sinica}			& 4.75 & 31.95 & 31.4 & 3.25 & 67.09 \\
& GumD.~\cite{chiu24_sinica}	        & 5.08 & 32.52 & \underline{30.3} & 5.64 & 62.59 \\
& HConv.		& \underline{4.47} & \underline{30.74} & 36.5 & \underline{2.88} & \underline{71.15} \\
\hline
\multirowcell{ 3}{Whspr  \\ \& Best \\ 2 SSLs } 
& WS.~\cite{chiu24_sinica}			& \underline{4.64} & 25.75 & 22.7 & 3.23 & 70.01\\
& GumD.~\cite{chiu24_sinica}     	& 4.80 & 27.54 & 23.4 & 5.52 & 63.71\\
& HConv.		& 4.74 & \underline{24.75} & \underline{22.4} & \underline{2.82} & \underline{71.71}\\


    \hline
  \end{tabular}
\end{table}

%% file: tables/largeModel_results.tex
\begin{table}[th]
  \caption{Evaluation results for large models across ASR and Non-ASR tasks.``LS'' denotes LibriSpeech, ``Mono'' indicates Mono-1h on ML-SUPERB. Their metrics are WER/CER. ``SV'' and ``ER'' stands for Speaker Verification and Emotion Recognition respectively and their metrics are Equal Error Rate and Accuracy. 
  For ``Best SSL'', we choose ``Data2Vec Large'' for LibriSpeech ASR and `WavLM Large'' for rest of them and for ``Whspr'', we choose``Whisper Medium''. 
  The interfaces are abbreviated as the following: ``WS.'' : ``Weighted Sum'' and  ``HConv.'': ``Hierarchical Convolution''.}
  \label{tab:largeModel_result}
  \small
  \centering
  \begin{tabular}{lccccc}
    \hline
    
      \multirowcell{ 2}{\textbf{Model}} & \multirowcell{ 2}{\textbf{Fusion} } & \multicolumn{2}{c}{\textbf{ASR}}  & \multicolumn{2}{c}{\textbf{Non-ASR}} \\
      \cline{3-6}
          &   & \textbf{LS}~$\downarrow$ & \textbf{Mono-1h}~$\downarrow$  & \textbf{SV}~$\downarrow$ & \textbf{ER}~$\uparrow$   \\
    \hline
  \hline
    \multicolumn{6}{c}{\textit{Single Model}} \\
    \hline


\multirowcell{ 2}{Best SSL} 
& WS.~\cite{yang21c_superb}			& 3.22 & 29.45 & 2.80 & 68.67  \\
& HConv.		& \underline{\textbf{3.10}} & \underline{28.26} & \underline{2.21} & \underline{72.49} \\
\hline
\multirowcell{ 2}{Whspr} 
& WS.~\cite{yang21c_superb}			& 5.59  & 24.78 & 3.76 & 71.48  \\
& HConv.		& \underline{5.25}  & \underline{23.36} & \underline{3.24} & \underline{72.96} \\
\hline
\hline
\multicolumn{6}{c}{\textit{2 Model Fusion}} \\
\hline
\multirowcell{ 2}{Whspr  \\ \& Best SSL } 
& WS.~\cite{chiu24_sinica}			& \underline{3.12} & 26.48 & 2.71 & 68.55 \\
& HConv.		& 3.55 & \underline{\textbf{22.87}} & \underline{\textbf{2.19}} & \underline{\textbf{73.52}} \\

    
    \hline
  \end{tabular}
\end{table}

%% file: sections/eval_setup.tex
\subsection{Tasks}
As stated in our motivation, we aim to comprehensively evaluate fusion by covering both ASR and non-ASR tasks.
For ASR, we use both the SUPERB~\cite{yang21c_superb} LibriSpeech ASR task as well as
the Monolingual and Multilingual ASR tasks from ML-SUPERB~\cite{shi23g_mlsuperb}.
In ML-SUPERB, ``Mono-1h'' is evaluated by training a model to conduct ASR on 13 languages respectively (each language 1h).
The CER for each of the 13 languages is averaged to produce an overall evaluation metric.
``Multi-1h'' on the other hand is the CER of doing Multilingual ASR simultaneously on 143 languages (each language 1h).
For non-ASR tasks, we select Speaker Verification~(SV) and Emotion Recognition~(ER) from SUPERB.
For SV, following~\cite{aldeneh24_SVDS}
, we simplify the default downstream module in SUPERB.
Although we were unable to reproduce the exact results from their paper, likely due to hardware differences, we observed improvements from simplifying the downstream model and adopted this architecture for all of our experiments.
For ER, we follow the default SUPERB setting.

\noindent
\subsection{Upstream Models}
To include both Self-supervised and Supervised Speech Models, we select 4 models: HuBERT Base, WavLM Base+, Data2Vec Base, and Whisper Small.
All of them have hidden states with 13 layers and hidden dimensions of 768.
Furthermore, to reduce the computation needed for fusion experiments, we constrain the combinations of these models.
For the SSL models, we investigate all possible combinations in 2 and 3 model fusion scenarios.
For SL + SSL fusion, we only select the best SSL model(s) to fuse with Whisper.
Notice that there are two options for Whisper Small, we could either use the multilingual version or the English-only version.
Through our preliminary experiments, we found out that the English version of Whisper does better than the multilingual version on LibriSpeech ASR.
Hence, we use the English version of Whisper for all tasks on LibriSpeech ASR and use the multilingual version for the rest of the tasks.

\subsection{Baselines and Proposed Interfaces}
According to~\cite{chiu24_sinica}, their best layer fusion + model fusion methods are either weighted sum~\cite{yang21c_superb} or dimension-wise Gumbel selection combined with temporal interleaving fusion.
For dimension-wise Gumbel selection, the authors used a Gumbel-softmax over a set of learned weights to determine which layer's dimension to select for the output.
Notice that the temporal interleaving fusion causes the input of the downstream model's time dimension to double, making it hard to compare them fairly under our framework.
Hence, we substitute temporal interleaving fusion as concatenation with projection (project back to the same dimension as the upstream models).
In our preliminary experiments on LibriSpeech ASR, we do not observe any difference after this revision.
In conclusion, we have two baselines in our experiments: either using weighted sum or dimension-wise Gumbel selection for layer fusion followed by concatenation and projection~\cite{chiu24_sinica}.
We denote them as ``WS.'' and ``GumD.'' respectively in our results.

For our proposed interfaces, we abbreviate the addition version as ``HConv.'' and the concatenation version as ``CHConv.''
Notice that CHConv.'s parameter count is much larger than HConv's due to the dense projection matrix after concatenation.
On the other hand, HConv. has roughly the same amount of trainable parameters as the baselines.
To make a fair comparison in terms of trainable parameters in our experiments, we regard HConv. as our main proposed interface and include it in all experiments.
For 2 model fusion experiments, we include CHConv. to demonstrate the potential improvement for HConv. if additional parameters are introduced.






%% file: sections/discussion.tex
\subsection{Comparison With and Without Fusion}
In Table~\ref{tab:main_result_single_col}, comparing the best results for single models and 2 model fusion, we observe that the performance with fusion is generally better than either of its constituent single models for all interfaces.
We compare the effectiveness of fusion by comparing the fusion performance with the best single model performance among the upstream models using the same interface. 
We find that the correlation across different interfaces is quite high, i.e. the fusion performance for all interfaces either improves over single models or they all under-perform the single models.
For instance, for Whisper Small and Data2Vec base fusion on Speaker Verification, we see a consistent improvement of $14.74\%$, $12.36\%$, $8.35\%$ for WS., HConv., and CHConv interfaces respectively. 
On the other hand, for Data2Vec Base and WavLM Base+ fusion on Speaker Verification, the improvements are $-1.2\%$, $-4.17\%$, $-9.49\%$ for WS., HConv., and CHConv interfaces respectively. 
These results suggest that selecting suitable upstream models for fusion is crucial and is more important than the specific selection of interface.


\subsection{Comparison between different Interfaces}
We find that in most cases, the proposed HConv. and CHConv. interfaces perform the best, especially for single models. For fusion, we observe the same trend except for Multilingual ASR on ML-SUPERB. For this task, the proposed methods approximately match the baseline methods.
We hypothesize the reason for this could be the language mismatch between the training and testing distributions for the upstream models.
Due to the fact that the SSL models we investigate are trained only on English, applying them to multilingual ASR is a more challenging task.

For the two baseline methods, we observe that WS. is more stable than GumD.
Particularly, for Speaker Verification tasks, using GumD. is significantly worse than WS.
We hypothesize that the speaker information is spread across several dimensions in each layer.
Hence, selecting different layers for each feature dimension 
compromises the integrity of the speaker information in the representation.
When comparing HConv. and CHConv., we see that CHConv. further improves the performance over HConv.,
especially for more challenging tasks such as Mono/Multi ASR or the fusion between Speech SL and SSL models.

\subsection{Fusion among Speech SSL and SL Models}
We also study the effect of fusion among SSL models vs fusion between SL and SSL models. 
We observe that fusing Whisper with other Speech SSL models often yields the best results over fusing only the Speech SSL models together. We hypothesize this is due to the fact that the SL and SSL models use very different training objectives, causing them to learn more heterogeneous representations. In contrast, all of the SSL models are trained using some form of masked prediction and therefore may learn more similar representations from model to model.
Also, the proposed interfaces HConv. and CHConv. consistently outperforms baseline interfaces, indicating that our interfaces are better at fusing between Speech SSL and SL models.
However, we find an exception on LibriSpeech.
In fact, Whisper by itself in the single model experiment is much worse than other Speech SSL models.
We hypothesize this is because the SSL models generally include LibriSpeech in their training set, whereas Whisper does not.

\subsection{Scalability for number of upstream models}
We further extend the number of our upstream models toward 3 models to test their scalability.
Surprisingly, we do not see an evident further improvement compared to 2 model fusion regardless of interface choices.
We hypothesize the reason for this could be that the representations from the upstream models are too similar and including more diverse models in the fusion could provide further improvements, but leave this investigation for future work.

\subsection{Scalability for Large Model Fusion}
Finally, we evaluate our fusion methods when scaling up model size.
To do this, we perform fusion with WavLM Large, Data2Vec Large, and Whisper Large (we also use the English version for LibriSpeech).
From Table~\ref{tab:largeModel_result}, we observe that the performance gap between our proposed interface and baseline is larger than that of the base models, which is consistent with the findings in~\cite{shih24_interface}.
Specifically, for large models, we have a gap of $3.61$  between WS. and HConv. for Mono-1h, and $4.97$ for Emotion Recognition.
On the other hand, we see a gap of $1.01$ between WS. and HConv. for Mono-1h, $0.71$ for Emotion Recognition when fusing the base models.
However, this trend is not consistent with Speaker Verification and LibriSpeech ASR.


%% file: sections/discussion_diff_langs.tex
\input{tables/mono1h_diff_lang}
To examine the effect of interface design with different Speech Foundation Models, we reports the performance per languages for 
Mono-1h experiment on ML-SUPERB.
The results for both base and large models are shown in Table~\ref{tab:mono1h_diff_lang}.

\subsection{Base Models}
From the single model experiments, we see that Whisper greatly outperforms the SSL models on non-English languages but lags behind for all three English datasets, which is reasonable due to the differences in dataset scale and distribution these models are trained on.
When switching the interface from Weighted Sum to Hierarchical Convolution, the improvements for SSL models are more consistent over all languages, while Whisper's improvements are marginal and mixed.

For the two model fusion experiments, Hierarchical Convolution Interface shows consistent improvements over two other baseline interfaces across all languages no matter which type of models are fused.
This demonstrates the robustness of our proposed methods when dealing with layer and model fusion together over baseline fusion methods.
Additionally, we observe that our method can effectively fuse two SSL models on non-English ASR tasks, while the other baseline interfaces cannot, i.e. 
On non-English tasks, ``WavLM \& Data2Vec'' get an average CER of $33.3$ when using WS. and  $31.8$ when using HConv.
Nonetheless, the best performance of single models for the each interface is WavLM with WS.: $33.0$ and WavLM with HConv. : $31.9$.

Comparing the difference between single and 2 model fusion, we see that ``WavLM \& Whisper'' performs the best by leveraging the strengths of WavLM model on English ASR and Whisper on non-English ASR.
Furthermore, even for some non-English ASR tasks, combining SSL and SL models gets a significant performance boost.
For example, the CER of ``fra2'' (using HConv.) for WavLM and Whisper are $42.8$ and $33.0$ respectively.
Fusion further improves this to $29.7$.

Lastly, we see that fusing all three models together does not get more improvement over the best 2 model fusion compared to the gap between single and 2 model fusion, i.e. ``WavLM \& Data2Vec \& Whisper''  outperform  ``WavLM \& Whisper'' by $0.13$ whereas ``WavLM \& Whisper'' outperform Whisper by $1.45$ on average.
The improvements mainly come from English tasks, which is reasonable since we fuse another SSL model (Data2Vec) that does better in English ASR.

\subsection{Large Models}
For single model experiments, WavLM Lg. is better on English tasks and worse on non-English tasks while Whisper Med. performs otherwise, which is the same trend as base models.
Comparing fusion and single model, we see that the performance gain of using HConv. comes from the English tasks.
Specifically, for non-English tasks, fusion gets the same average CER ($23.3$) as Whisper Med., while the average CER of English tasks goes from $23.77$ to $21.53$.
On the other hand, when using WS., the non-English average CER of fusion falls between the performance of single models.
This result shows the advantage of choosing HConv. as the fusion interface, especially for low-resource languages.

For the 2 model fusion experiments, we see a large gap between using WS. and HConv. for large models compared to the base models consistent across all languages.
For ``WavLM \& Whisper'', the average CER gaps between WS. and HConv. are $0.33$ and  $1.3$ for English and non-English respectively.
In contrast, ``WavLM Lg. \& Whisper Med.'' achieve a improvement of $3.33$ and $3.90$.
These results emphasize our proposed method's scalability on large SFMs, especially for low-resource languages.

%% file: tables/mono1h_diff_lang.tex
\begin{table*}[h!]
    \setlength{\tabcolsep}{4pt}
  \caption{ASR Evaluation results for different interfaces and model combination across different languages. The metrics in this table are WER/CER, the lower the better. 
  For the base upstream models, ``WavLM'' and  ``Whisper'' denotes ``WavLM Base+'' and ``Whisper Small''. 
  For the large upstream models, ``WavLM Lg.'' and ``Whisper Med,'' infers ``WavLM large'' and ``Whisper Medium'' respectively.
  The interfaces are abbreviated as the following: ``WS.'' : ``Weighted Sum'', ``HConv.'': ``Hierarchical Convolution'' and ``GumD.'': ``Gumbel Dimension Selection''.}
  \label{tab:mono1h_diff_lang}
\normalsize
  \centering
  \begin{tabular}{cccccccccccccccc}
    \hline
    
      \multirowcell{ 2}{\textbf{Model}} &  \multirowcell{ 2}{\textbf{Interface}} & 
      \multirowcell{ 2}{\textbf{Avg.}} & \multicolumn{13}{c}{\textbf{Languages}} \\
      \cline{4-16}
      & & &	eng1 &	eng2 &	eng3 &	fra1 &	fra2 &	deu1 &	deu2 &	rus &	swa &	swe &	jpn &	cmn &	xty \\
    \hline
\hline
\multicolumn{16}{l}{\textbf{Base Model:}}\\
\multicolumn{16}{c}{\textit{Single Model}} \\
\hline
\multirowcell{ 2}{WavLM} 
& WS.~\cite{yang21c_superb}			& 32.01 &	23.4 &	32.5 &	26.7 &	38.2 &	41.9 &	27.0 &	32.0 &	28.8 &	28.7 &	26.3 &	13.8 &	32.2 &	61.2 \\
& HConv.		& 30.74 &	21.5 &	30.5 &	27.6 &	36.8 &	42.8 &	27.0 &	31.7 &	28.4 &	23.9 &	25.6 &	13.5 &	31.0 &	58.6 \\
\hline
\multirowcell{ 2}{Whisper} 
& WS.~\cite{yang21c_superb}			& 26.27 &	25.1 &	34.6 &	28.6 &	30.2 &	31.8 &	20.4 &	25.4 &	18.1 &	26.0 &	20.1 &	9.3 &	18.2 &	61.4 \\
& HConv.		& 26.23 &	25.8 &	32.5 &	31.1 &	28.8 &	33.0 &	18.6 &	26.9 &	17.8 &	21.3 &	20.1 &	9.1 &	18.3 &	66.0 \\
\hline
\multirowcell{ 2}{Data2Vec} 
& WS.~\cite{yang21c_superb}			& 35.77 &	19.7 &	32.5 &	26.0 &	41.8 &	45.4 &	31.0 &	36.3 &	33.1 &	32.0 &	30.2 &	17.2 &	38.3 &	67.8 \\
& HConv.		& 34.22 &	19.0 &	30.5 &	24.1 &	41.2 &	44.7 &	28.3 &	36.5 &	32.1 &	28.0 &	29.0 &	16.6 &	38.3 &	64.1 \\
\hline
\hline
\multicolumn{16}{c}{\textit{2 Model Fusion}} \\
\hline
\multirowcell{ 3}{WavLM  \\ \& Data2Vec } 
& WS.~\cite{chiu24_sinica}			& 32.02 &	19.7 &	30.9 &	25.8 &	38.7 &	41.9 &	27.9 &	31.5 &	28.8 &	26.5 &	26.8 &	14.1 &	32.6 &	63.9 \\
& GumD.~\cite{chiu24_sinica}         & 32.93	&   20.9 &	32.6 &	25.8 &	39.1 &	41.2 &	28.2 &	33.2 &	30.2 &	28.4 &	27.7 &	14.2 &	34.2 &	64.4 \\
& HConv.		& 30.62 &	18.7 &	30.1 &	23.4 &	36.4 &	39.9 &	26.1 &	31.3 &	27.9 &	23.3 &	26.0 &	13.7 &	31.6 &	62.2 \\
\hline
\multirowcell{ 3}{WavLM \\  \& Whisper } 
& WS.~\cite{chiu24_sinica}			& 25.89 &	23.3 &	31.5 &	27.7 &	30.1 &	33.0 &	20.1 &	25.1 &	18.2 &	24.2 &	19.9 &	9.7 &	18.9 &	60.5 \\
& GumD.~\cite{chiu24_sinica}         & 27.37 &	24.6 &	34.3 &	28.9 &	31.8 &	33.8 &	21.7 &	26.8 &	20.0 &	24.1 &	22.5 &	10.0 &	21.2 &	62.2 \\
& HConv.		& 24.88 &	23.0 &	31.1 &	27.4 &	27.4 &	29.7 &	18.3 &	24.5 &	17.1 &	20.3 &	19.2 &	9.2 &	18.2 &	62.8 \\
\hline
\hline
\multicolumn{16}{c}{\textit{3 Model Fusion}} \\
\hline
\multirowcell{ 3}{WavLM   \& Data2Vec\\ \& Whisper } 
& WS.~\cite{chiu24_sinica}     & 25.75 &	20.5 &	29.7 &	25.6 &	31.0 &	32.4 &	20.7 &	25.1 &	18.8 &	24.1 &	20.5 &	9.7 &	19.8 &	59.0 \\
& GumD.~\cite{chiu24_sinica}   & 27.54 &	21.1 &	31.8 &	25.3 &	32.3 &	36.1 &	22.4 &	28.1 &	21.0 &	24.5 &	22.8 &	10.0 &	21.5 &	62.5 \\
& HConv.  & 24.75 &	22.1 &	30.2 &	26.6 &	28.0 &	30.6 &	18.2 &	25.5 &	17.1 &	20.3 &	19.2 &	9.3 &	18.0 &	61.4 \\

\hline
\hline
\multicolumn{16}{l}{\textbf{Large Model:}}\\
\multicolumn{16}{c}{\textit{Single Model}} \\
\hline
\multirowcell{ 2}{WavLM Lg.} 
& WS.~\cite{yang21c_superb}			& 29.45 &  21.2 &	29.7 &	25.0 &	33.3 &	39.8 &	24.2 &	29.8 &	25.9 &	26.2 &	24.4 &	11.8 &	28.6 &	59.3 \\
& HConv.		& 28.26 &  20.1 &	26.7 &	24.5 &	32.3 &	40.2 &	23.7 &	29.3 &	26.5 &	20.0 &	23.7 &	11.9 &	27.2 &	58.5 \\
\hline
\multirowcell{ 2}{Whisper Med.} 
& WS.~\cite{yang21c_superb}			& 24.78 & 22.7 &	31.6 &	29.1 &	28.5 &	30.9 &	20.7 &	23.8 &	15.0 &	23.8 &	17.4 &	8.3 &	16.8 & 62.0 \\
& HConv.		& 23.36 & 21.9 &	28.7 &	27.4 &	24.8 &	32.0 &	16.4 &	25.2 &	13.9 &	19.1 &	16.3 &	8.5 &	16.5 & 60.7 \\
\hline
\hline
\multicolumn{16}{c}{\textit{2 Model Fusion}} \\
\hline
\multirowcell{ 2}{WavLM Lg. \\  \& Whisper Med. } 
& WS.~\cite{yang21c_superb}			& 26.48 &	20.7 &	29.2 &	24.7 &	31.1 &	36.6 &	21.8 &	26.6 &	19.8 &	22.8 &	21.2 &	10.6 &	21.6 &	59.4 \\
& HConv.		& 22.87 &	18.4 &	24.3 &	21.9 &	25.6 &	31.1 &	17.3 &	22.4 &	15.1 &	18.9 &	16.6 &	9.4  &	19.0 &	57.1 \\
\hline
  \end{tabular}
  \vspace{-6pt}
\end{table*}

%% file: sections/conclusion.tex
In this work, we propose a unified framework for model and layer fusion of speech foundation models and see that our proposed fusion method experimentally leads to consistent performance improvements across multiple speech tasks.  
Furthermore, from the evaluation results, our takeaway is that selecting the right Speech Foundation Models to fuse is important in the first place. 
We also observe that fusing the supervised Whisper model with self-supervised models consistently outperforms the fusion of multiple self-supervised models, indicating that supervised and self-supervised learning objectives can be synergistic.
Although our work show that fusing multiple upstream models results in improved performance over each constituent model, the inference computational cost scales up accordingly. Thus, a promising avenue for future work is to explore joint distillation of multiple upstream models as an alternative to fusion.